\documentclass{aa}
\usepackage{latexsym, epsfig}
\usepackage{graphicx}
\usepackage{natbib}

\def\kms{kms$^{-1}$}

\begin{document}

\renewcommand{\thefootnote}{\fnsymbol{footnote}}

\title{Probing Turbulence with Infrared Observations in OMC1}

\author{M.Gustafsson\inst{1}\thanks{Based on observations obtained at the
    Canada-France-Hawaii Telescope (CFHT) which is operated by the
    National Research Council of Canada, the Institut National des
    Science de l'Univers of the Centre National de la Recherche
    Scientifique of France, 
and the University of Hawaii.}\and D.Field\inst{1}\and
  J.L.Lemaire\inst{2}\and F.P.Pijpers\inst{3}}

\institute{Department of Physics and Astronomy, University of Aarhus,
  DK-8000 Aarhus C, Denmark \\ \email{maikeng@phys.au.dk}\and  Observatoire de Paris \&
  Universit\'e de Cergy-Pontoise, LERMA \& UMR 
8112 du CNRS, 92195 Meudon, France\and Space and Atmospheric
  Physics, Dept. Physics, Imperial College London, England}

\date{For Main Journal, Diffuse Matter in Space. Received:  Accepted:} 

\abstract{ A statistical analysis is presented of the turbulent
  velocity structure in the Orion Molecular Cloud at scales ranging
  from 70 AU to 3$\cdot10^4$ AU. Results are based on IR Fabry-Perot
  interferometric observations of shock and photon-excited H$_2$ in
  the K-band S(1) v=1-0 line at 2.121$\mu$m and refer to the dynamical
  characteristics of warm perturbed gas. Data consist of a spatially
  resolved image 
  with a measured velocity for each resolution limited region
  (70AUx70AU) in the image. The effect of removal of apparent large
  scale velocity gradients is discussed and the conclusion drawn that
  these apparent gradients represent part of the turbulent cascade and
  should remain within the data. Using our full data set, observations
  establish that the Larson size-linewidth relation is obeyed to the
  smallest scales studied here extending the range of validity of this relationship by nearly 2
  orders of magnitude. The velocity probability distribution function
  (PDF) is constructed showing extended exponential wings, providing evidence of intermittency, further supported by
  the skewness (third moment) and kurtosis (fourth
  moment) of the velocity distribution. Variance and kurtosis of the
  PDF of velocity differences are constructed as a function of
  lag. The variance shows an
  approximate power law dependence on lag, with exponent significantly lower
  than the Kolmogorov value, and with deviations below 
  2000AU which are attributed to outflows and possibly disk structures associated with low mass star
  formation within OMC1. The kurtosis shows strong deviation from a
  gaussian velocity field, providing evidence of velocity correlations at small
  lags. Results agree accurately with semi-empirical simulations
  in~\cite{eggers1998}.

In addition, 170 individual H$_2$ emitting clumps have been
analysed with sizes between 500
and 2200 AU. These show considerable diversity with regard to PDFs and variance functions (related to second order structure functions) displaying a variety of shapes of the PDF and
different values of the scaling exponent within a restricted spatial
region. However, a region associated with an outflow from a deeply
embedded O-star
shows high values of the scaling exponent of the variance
function, representing a strong segregation of high and low exponent
clumps. Our analysis constitutes the
first characterization of the turbulent velocity field at the scale of
star formation and provide a dataset which models of star-forming
regions should aim to reproduce.

}

\maketitle

\markboth{M. Gustafsson et al: }{Probing Turbulence with Infrared Observations in OMC1}

\section{Introduction}

A key to understanding the mechanism of star formation is to
characterise in detail the nature of the turbulent, weakly ionized and
magnetised plasma in which stars form. Recently Gustafsson et al. 2003
(Paper I) published observational results for the vibrational emission
of H$_2$ in the archetypal star-forming region in the Orion Molecular
Cloud, OMC1. Using a combination of Fabry-Perot interferometry and the
PUEO adaptive optics system on the Canada-France-Hawaii Telescope
(CFHT), data achieved a spatial resolution of 0\farcs15~(70AU at the
distance of OMC1, D=460 pc \citep{bally}), with a velocity discrimination of 1
kms$^{-1}$ in regions of high brightness. These data, limited as they
are to highly excited regions in OMC1, provide the first
opportunity to study the physical properties of a star-forming region
both in terms of its morphology and bulk gas motion at the scale of
star formation.  
The aim of the present paper is to characterise the nature of the
turbulent velocity field in OMC1 for the subset of regions represented
through vibrationally excited H$_2$.  Our results should help to
provide a benchmark for comparison with MHD models of star-forming
regions.

The importance of turbulent gas motion in regulating star formation
may be illustrated as follows. Relative bulk motions obtained from observations reported here range up to 40 kms$^{-1}$ in dense gas. Since these motions are supersonic, it is evident that bulk motion must contribute more energy per unit
volume, that is, pressure, than thermal energy. Equally if a simple
scaling law is used between magnetic induction, B, and particle
density, n, whereby B=b$n^{1/2}\mu$G, with n in
cm$^{-3}$ \citep{troland1986}, then the energy contained in  
bulk mass motion will exceed the energy in the magnetic field, B$^2$/2$\mu$ (in SI units), for
bulk velocities greater than 2.2 b kms$^{-1}$. The constant b
lies typically between 1 and 2 \citep{kristensen2005}. Thus in many
regions the 
turbulent pressure may exceed the magnetic pressure, given that bulk motion proceeds at tens of \kms as mentioned above. The magnetic
pressure in turn typically exceeds the thermal pressure, the ratio of
thermal to magnetic pressure being $\sim$5x10$^{-3}$ T/b$^2$, where T is
the gas temperature. In the above simple prescription, any anisotropy
of the magnetic field has been ignored. At all events, on
purely energetic grounds, in earlier stages of star  formation,
turbulence is the controlling support mechanism against gravitational
collapse, if bulk mass motions exceed a few kms$^{-1}$. This statement remains true even in the warm regions involved in the present study.

Observational and computational evidence for the simple picture
presented above has grown considerably in the last decade as described in recent reviews by \cite{larson2003} and
\cite{maclow2004}. Moreover the role of turbulence goes further than
simply 
slowing the process of gravitational collapse. Simulations show that
turbulence may determine the initial mass function (IMF) for star
formation through turbulent fragmentation \citep[e.g.][]{nordlund2003}. It is evident that
MHD models, which make such far-reaching predictions, need to be
constrained by observations. The meeting of theory and observation may
be achieved by recording the statistical properties of velocity fields
and this standard approach is adopted here. 
In the present work we make only very limited comparison with
published models. Rather, we present modellers with characteristic
behaviour which it is their challenge to reproduce.

Several techniques have previously been used to characterize the
structure of brightness and velocity in molecular clouds based largely upon CO observations, reviewed by \cite{goodman1998} and
\cite{miesch1999}. These earlier observations, tracing relatively cool and low density gas, are limited in spatial
resolution and can only be used to address the physics at scales larger 
than roughly 0.03pc (6000AU). When dealing with turbulence it is assumed that
most features are scale-free between the driving and dissipation
scale and that the behaviour for example of the structure functions
(see Sect.~\ref{sec:2sf_full}) 
can be extrapolated to the very much smaller scales of individual star
formation. This is however far from certain. The high spatial
resolution infrared data reported in Paper I give the opportunity to
discover how structure functions develop at much smaller scales than
have hitherto been probed.

In this paper we use three standard statistical measures to provide observational
constraints on theories and simulations concerning gravitational
collapse and star formation in a turbulent ISM. First, we test whether
the observed scaling behaviour of the 
velocity structure, encapsulated in the size-linewidth relation or "Larson law"
\citep{larson1981}, holds for the smaller scales involved
here (Sect.~\ref{sec:larson}). Second, we use probability distribution
functions (PDFs) of 
velocities (Sect.~\ref{sec:pdf_full}) as a probe of intermittency
\citep{falgarone1990,falgarone1991,falgarone1994,miesch1995,miesch1999,ossenkopf02}.
Third, we calculate the low order moments 
of the velocity difference PDF
\citep{miesch1995,lis1998,miesch1999,ossenkopf02} as a function of lag
(Sect.~\ref{sec:2sf_full}), that is, the structure functions or
functions directly related to structure functions. In addition, the
high spatial resolution of the 
data allows us to go further than determining these quantities for
the observed field as a whole. In Sect.~\ref{sec:clump} we
identify 170 individual clumps in OMC1 with a mean size of
1300 AU. We
calculate the individual PDFs of peak velocities and the variance functions 
for these clumps.

\section{Observations and data reduction}
\label{sec:data}

\begin{figure}
\resizebox{\hsize}{!}{\includegraphics{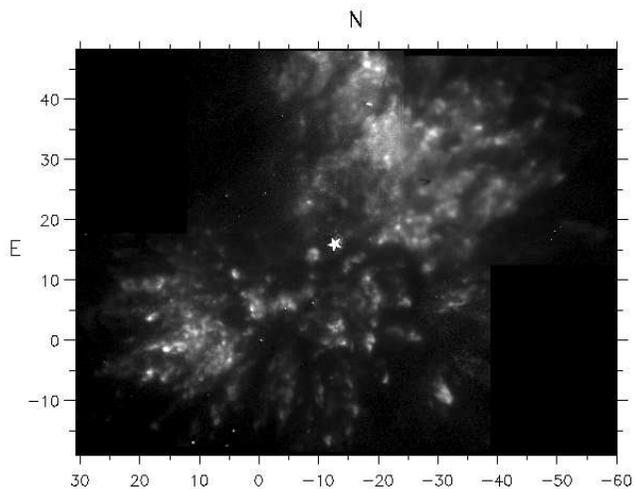}}
\caption{\small Velocity integrated emission in the S(1) v= 1-0 H$_2$
  emission line covering the full observed field in OMC1. (0,0) marks
  the position of TCC0016 05$^{\mathrm{h}}$35$^{\mathrm{m}}$14\fs91,
  -05\degr22'39\farcs31 (J2000). The star marks the position of BN. Units are arcseconds, where 1 arcsecond = 460AU.}
\label{fig:field}
\end{figure}

Near infrared K-band observations of the strongest H$_2$ emission line in
Orion, v= 1-0 S(1) at 2.121 $\mu$m, are used as a tracer of radial gas velocity to provide the data for the statistical analysis performed here.
The observational data are the same as those described in
Paper I and a detailed
description of the data acquisition and reduction may be found there.
In brief, OMC1 was observed using the CFHT with GriF on the night of
December 5th 
2000. The GriF instrument \citep{clenet}, is a combination of the
PUEO adaptive optics (AO)  
system on the CFHT with interferometric spectral scanning. The
interferometer, a Queensgate ET50WF Fabry-Perot (FP), affords a
measured spectral resolution $\lambda$/$\Delta$$\lambda$$\sim$~2030,
that is, 150 kms$^{-1}$. The detector has a field of view of 36\arcsec
$\times$ 36\arcsec with a pixel scale of 0\farcs035. The region of
OMC1 observed, shown in Fig.~\ref{fig:field}, consists of four
overlapping fields 
and the
entire region is centred 15\arcsec N and 15\arcsec W of TCC0016
(05$^h35^m14.\!^s91$, 
  $-05^{\circ}22^{\prime}39\farcs31$, J2000).

Each field has been scanned through the H$_2$ line from a wavelength
in the far blue wing to a wavelength in the far red wing, using
step sizes between 4.4 $\times$ 10$^{-4} \mu$m and $4.6\times10^{-4}\mu$m
($\sim 65$\kms), allowing adequate sampling of the
instrumental profile. To
prevent superposition of different FP orders 
during scanning, a H$_2$ v=1-0 S(1) interference filter with a
2.122$\mu$m central wavelength and a bandwidth of 0.02$\mu$m, was
inserted between the FP and the detector.  For each region and
 each wavelength, a single exposure of 400s was performed. 
Data reduction includes dark and bias subtraction, 
flat-fielding,  bad pixel rejection, image
recentering, 2D wavelength correction and
subtraction of sky background obtained from an image in the far wing of
the H$_2$ profile. 
Various reference stars for the AO were used: TCC0016
(m$_V \sim$ 14) in the SE and SW field, Parenago 1838 (m$_V
\sim$ 8) in NE and Parenago 1819 (m$_V \sim$ 12) in NW. The FWHM of
the point-spread-function of stars in the region has been examined yielding
a spatial 
resolution of 0\farcs15. This near diffraction limited spatial
resolution was maintained throughout all observations.
In view of the spatial resolution of 0\farcs15, all images have been smoothed by a moving average of 3 by 3 pixels to improve S/N without significantly degrading the spatial resolution. 

Radial velocities were found by choosing a specific position on the
sky and taking a cut through the cube made up of the channel maps. This yielded a set of count rates as a function of wavelength
constituting a line profile and a Lorentzian is fitted to the profile \citep{clenet}. The position of the 
peak in emission gives the radial 
velocity. The Lorentzian form was chosen because it represents
  the instrumental profile of the Fabry-Perot. We note that the
  instrument profile is highly symmetric \citep{clenet} and thus a
  choice for example of a centroid rather than the peak of a
  Lorentzian is immaterial to an estimation of the velocity. We also note that
  less than 2\% of profiles are double peaked: the velocity of the
  brighter of the two peaks is chosen when these cases are encountered
  (Paper I).

Performing these operations pixel by pixel for the full field
of view results in very accurate fits for the brighter regions. Including
statistical error in the count rates, relative peak wavelengths may be
found in bright local zones to 2 to 3
parts per 10$^6$$ (3\sigma)$, that is, to better than
$\pm$1kms$^{-1}$. The quality of the fit decreases with decreasing
brightness. Since an assessment of errors is important in
discussing the statistical properties of the velocity field, the
accuracy of determination of the central wavelength has been
estimated by performing a large number of fits for a range of pixel
brightness. An empirical 
correspondence between brightness and error in velocity has thereby
been determined. This may be expressed as follows:  
\begin{eqnarray}
\sigma&=&0.15+18.1\exp(-I/4.8\cdot 10^{-2})\nonumber\\
&+&2.15\exp(-I/1.2))
\label{eqn:errors}
\end{eqnarray}
where $\sigma$ is the standard deviation of the radial velocity in \kms ~and I
is the brightness in counts $\cdot$ pixel$^{-1}\cdot s^{-1}$ for the peak emission associated with the Lorentzian fit. All pixels
with brightness below 0.05 counts $\cdot$ pixel$^{-1}\cdot s^{-1}$, that is 
2\% of the maximum brightness, have
been excluded. Thus the largest uncertainty, $\sigma$, in the radial
velocity that may be encountered in our data is 8.6 \kms. 

In addition to random errors in the velocity, systematic
  errors may also occur in establishing velocity differences between
  regions which are physically remote. This may take place due to distortions in the 2D
  wavelength correction 
  arising from possible mechanical instabilities or drift. In this connection, observations of a
  restricted part of the full field comprising two
  36\arcsec$\times$36\arcsec fields to the SE and NW were performed in
  January 2003. Comparisons have been made of the velocity fields in
  the 2000 data, used here, and in the 2003 data. Average velocities
  have been estimated within regions of about 50$\times$50 pixels
  (1\farcs75$\times$1\farcs75) separated by up to 60\arcsec, the full
  size of the image. We find that velocity differences in the 2000
  data and the 2003 data agree to better than $\pm$1\kms. Since these
  results were obtained using independent calibration data, they provide
  a clear indication that systematic errors in velocity differences between remote regions are not present.

Conversion of count rates to
absolute values of surface brightness was obtained by comparison with
calibrated data in \cite{vannier}. Thus the count rate
of 2.15 counts $\cdot$ pixel$^{-1}\cdot s^{-1}$, at the peak of emission to the SE
of TCC0016, corresponds to a velocity integrated brightness of
3.0$\pm0.15 \cdot 10^{-5}$~Wm$^{-2}$sr$^{-1}$.

The composite field observed within OMC1, shown in
Fig.~\ref{fig:field}, is 90\arcsec 
$\times$ 70\arcsec. This allows us to address statistical
issues involving scales ranging from 70 to 3$\cdot 10^4$ AU or 3.4$\cdot10^{-4}$ to 0.15 pc.

Velocities are expressed as $v_{lsr}$
by assigning a mean velocity of all the data of 12$\pm$6~\kms ~in the
local standard of rest
\citep{chrysostomou1997,salas,odell2001}. Since
we are essentially concerned only with the relative velocities of
regions within OMC1, the uncertainty in the value of $v_{lsr}$ is not
material to our discussion.   

\section{H$_2$ emission and excitation in the inner region of OMC1}
\label{sec:excitation}
There is a wealth of data for the dense star-forming region of OMC1, even if consideration is
restricted to IR observations of H$_2$ emission. Reference to these
extensive data, covering features such as the well-known
"bullets" and  "fingers" of H$_2$ emission, and very high
spatial resolution observations of H$_2$ performed using the HST, VLT
and CFHT may be found in recent work  e.g.  \cite{mccaughrean1997};
\cite{stolovy}; \cite{schultz}; \cite{lee}; 
\cite{odell2001}; \cite{davis2001}; 
 \cite{vannier}; \cite{doi2002}; \cite{gustafsson2003} (Paper I);
 \cite{lacombe2004}. Our 
account here is restricted to data specifically concerned with radial
motion measured in Orion using H$_2$ IR emission as a
tracer \citep{sugai,chrysostomou1997,salas}. Excitation mechanisms of H$_2$ are however briefly considered here since these have a bearing on later discussion. 

The high brightness of H$_2$ emission in the v=1-0 S(1) line,
frequently exceeding 10$^{-5}$Wm$^{-2}$sr$^{-1}$, indicates that the
gas is of high column density. Since brightly emitting clumps are
typically 1$\arcsec$ to 2$\arcsec$ in angular size (1$\arcsec$ =
2.2$\cdot$10$^{-3}$pc), the inference is that number densities are also
high. The most feasible general mechanism of H$_2$ excitation, for high
brightness, is through C-type (magnetic) shocks \citep{smith1990,kaufmanb,kaufmana,timmerman,wilgenbus,vannier,bourlot,flower2003,kristensen2003,kristensen2005}. Low
velocity shock excitation in dense gas in regions of OMC1 
shown in Fig.~\ref{fig:field}, has been analysed in detail for example in
\cite{vannier} and in \cite{kristensen2003}. Densities in the
post-shock gas exceed 10$^7$cm$^{-3}$. In addition, the data of Paper
I graphically illustrate the presence of shocks through
the clear spatial association of bright H$_2$ emission with gas flows.  

In addition to shocks, photon induced excitation of H$_2$, in so-called
photon dominated regions (PDRs), can play a role in yielding the
observed H$_2$ IR emission. This has been discussed in detail in 
\cite{kristensen2003} who show that PDRs may contribute significantly
in specific regions, for example in the SE (Peak 2) region, especially at the edge
of clumps of material in this region.  In strongly emitting zones,
PDRs cannot yield sufficient brightness, 
even including effects of advection \citep{lemaire1999}, high density, and the high UV flux
from $\Theta$$^1$Ori-C, in excess of 10$^5$ times the standard UV field \citep{storzer}.

\section{Statistical Analysis}
\label{sec:stat}
\subsection{Filtering of data}
\label{sec:filter}

The presence of large scale velocity differences in OMC1 has been
known for many years. \cite{sugai}, \cite{chrysostomou1997}
and~\cite{salas} have all used FP interferometry in OMC1 in the IR and
studied the large scale velocity structure of the hot H$_2$
component. From our observations (Paper I) we find that the gas in
Peak 1 (NW 
of BN) is on average 10 \kms ~more blueshifted than in Peak 2 (SE of BN: Fig.~\ref{fig:field}; see also Fig.~2 in
Paper I). This velocity 
difference has generally been viewed in the context of an outflow
mechanism from the IRc2-complex and has previously been explained 
by an expanding shell \citep{scoville,sugai}, a bipolar
outflow \citep{chrysostomou1997} or a spherical wind
\citep{salas}. 
The velocity difference could however be a manifestation of turbulent
energy injected into the system and stirring of the gas in large scale
vortices. Thus the nature of 
the velocity difference remains uncertain and in the following we
will explore this question further.

As discussed in \cite{miesch1994}, \cite{miesch1995} and
\cite{miesch1999}, large scale trends - such as a general velocity gradient - could dominate any
statistical quantifier and give misleading results. This is due to the
fact that the
statistical methods lose spatial information and therefore
depend on an absence of systematic trends within the data. 
It may therefore be argued that any large scale trends should
be removed and only local velocity fluctuations
studied. \cite{ossenkopf02} however point out that large scale systematic motions
may be part of the turbulent cascade if these inject energy into the
system and should therefore only be removed if the turbulence studied
is exclusively driven on smaller scales. 

In this connection, observations of molecular clouds seem to indicate that interstellar
turbulence is driven on scales larger than the clouds themselves, and
arguments have been presented that supernovae explosions dominate
\citep{maclow2004}. However, contributions to the continuous driving of
the turbulence may also arise from protostellar outflows
\citep{matzner2000} and stellar winds from massive O stars
\citep{maclow2004}. OMC1 is itself a region of highly active star formation. Observations ranging
from X-ray \citep{garmire,feigelson2002} to radio
wavelengths \citep[e.g.][]{churchwell1987,zapata2004} reveal the 
presence of protostars, 
outflows \citep[e.g.][]{genzel}, HH objects \citep[e.g.][]{odell1997b,doi2004} and mainline OH, H$_2$O and SiO maser emission
\citep[e.g.][]{norris,menten1995,gaume1998,greenhill2004b}, noting that maser emission of the nature observed in OMC1 is associated with the presence of O-stars.  
Thus, in OMC1 the turbulence could be driven locally by the powerful outflow
from the BN-IRc2 
complex, which contains a number of possible candidates such as source
I, source n and BN itself
\citep{menten1995,gezari,doeleman1999,greenhill2004a,shuping2004}.
Turbulence may also be driven 
on smaller scales by low mass star formation.

The alternative mechanism remains, as mentioned earlier, that the larger
scale dynamics is predominantly
part of a turbulent cascade driven on greater scales. For this
reason we test the statistical 
quantifiers explored in this paper with the full velocity field
and also the residual velocity fluctuations resulting from removing the
large scale velocity trends.

The procedure of removal of large scale gradients, or filtering, follows
\cite{miesch1994}. The removal is performed by convolving the data with a smoothing function and then subtracting the
smoothed image from the original to obtain the residual image of the
small scale
velocity fluctuations. The residual image should then reveal any small
scale structures that were superposed on the large scale trends in the
original. The filtering or smoothing function used here is a 
two-dimensional equally weighted moving average in the form of a
square box. The 
optimal size of the filter is the broadest possible that removes the
large scale gradient but leaves all other features. The 2D autocorrelation
function of a velocity map can be used for detecting velocity gradients. These show up as anticorrelations in the direction of the
gradient and recorrelations in the direction at right angles \citep{spicker}. 

In the autocorrelation function of our velocity map
the correlation persists in the NE-SW direction and decorrelation
occur in the NW-SE direction. This indicates a large scale velocity
gradient in the NW-SE direction which is consistent with the velocity
difference between Peak 1 and 2. The widest 
filter where no anticorrelated sidelobes are discernible in the
autocorrelation function of the residual image is the optimal. By calculating the autocorrelation function for filters of varying size,
the optimal filter size was found to be 14\arcsec $\times$
14\arcsec (6400 $\times$ 6400 AU). All further analysis has been carried out on both the
original full velocity field and on the residual image obtained from filtering with the optimal filter. This latter will be
referred to as the filtered velocity image or field. We find that it is certainly unwise to remove apparent large scale
velocity trends and that the filtered image is not representative of the dynamics in OMC1, as discussed immediately below in Sect.~\ref{sec:larson}.

\subsection{Size-linewidth relation}
\label{sec:larson}

\cite{larson1981} identified an empirical relation between
linewidth (or velocity dispersion) and size for molecular clouds and clumps within clouds over
scales of 
$\sim$ 0.1-100 pc. Larson obtained a power law
\begin{eqnarray}
 \Delta v_{obs} \propto R^{\alpha} \;\;\;\;\;  \alpha \simeq 0.38
\label{eqn:larson}
\end{eqnarray}
where $\Delta v_{obs}$ is the observed linewidth and R is the
effective radius of the object. This relationship essentially supports a model in which larger motions are associated with larger scales.  
More recent studies give a range of values for
$\alpha$ over dimensions of 0.02-100 pc \citep[e.g.][]{caselli1995,peng1998} showing $\alpha$ varying between 0.2 and 0.7. Some 
studies indicate that massive 
star forming regions tend toward the lower values
\citep[e.g.][]{caselli1995}. There are however indications that the
relation in eq. (2) breaks down in the most massive star forming regions
\citep{plume1997}. A comprehensive overview was given in
\cite{goodman1998}.  
Size-linewidth measurements require that objects are well defined
within a map and that a linewidth and size can be objectively assigned to objects within the field of observation.
The definition of objects is obvious for isolated clouds, but
encounters a certain degree of arbitrariness when dealing with clumps
in a cloud. 

The structure of OMC1, measured in vibrationally excited H$_2$, reveals a very clumpy environment where the regions of
bright emission are in many circumstances imposed on extended weaker emission. OMC1 is thus a case study in the difficulty
of determining the physical extent of individual clumps in a turbulent
environment. In order 
to circumvent this problem in computing a size-linewidth relation we
follow the method developed by \cite{ossenkopf02}. We remind the
reader at this stage that our observations of linewidth are
indirect. That is, we obtain the velocity associated with any pixel
from the peak of the line profile observed with the low resolution
Fabry-Perot, see Sect.~\ref{sec:data}. Thus the linewidth associated
with any chosen assembly of pixels is given by the velocity
dispersion of these peak velocities. This is in contrast to
the standard technique in radio observations in which the velocity
resolution of observations is a fraction of the linewidth, and the
linewidth may then be obtained directly from the measured lineshape.
The 
technique which we use here, that of taking the peak of the line for
each pixel, was also used by~\cite{ossenkopf02} (using centroid
velocities). For this reason we refer below to the Larson relationship as a
size-velocity dispersion relation, rather than the more familiar
size-linewidth. In the present case 
we note that H$_2$ emission is optically thin and there are no optical
depth effects, such as those discussed in detail in~\cite{ossenkopf02}.  

\begin{figure}
\resizebox{\hsize}{!}{\includegraphics{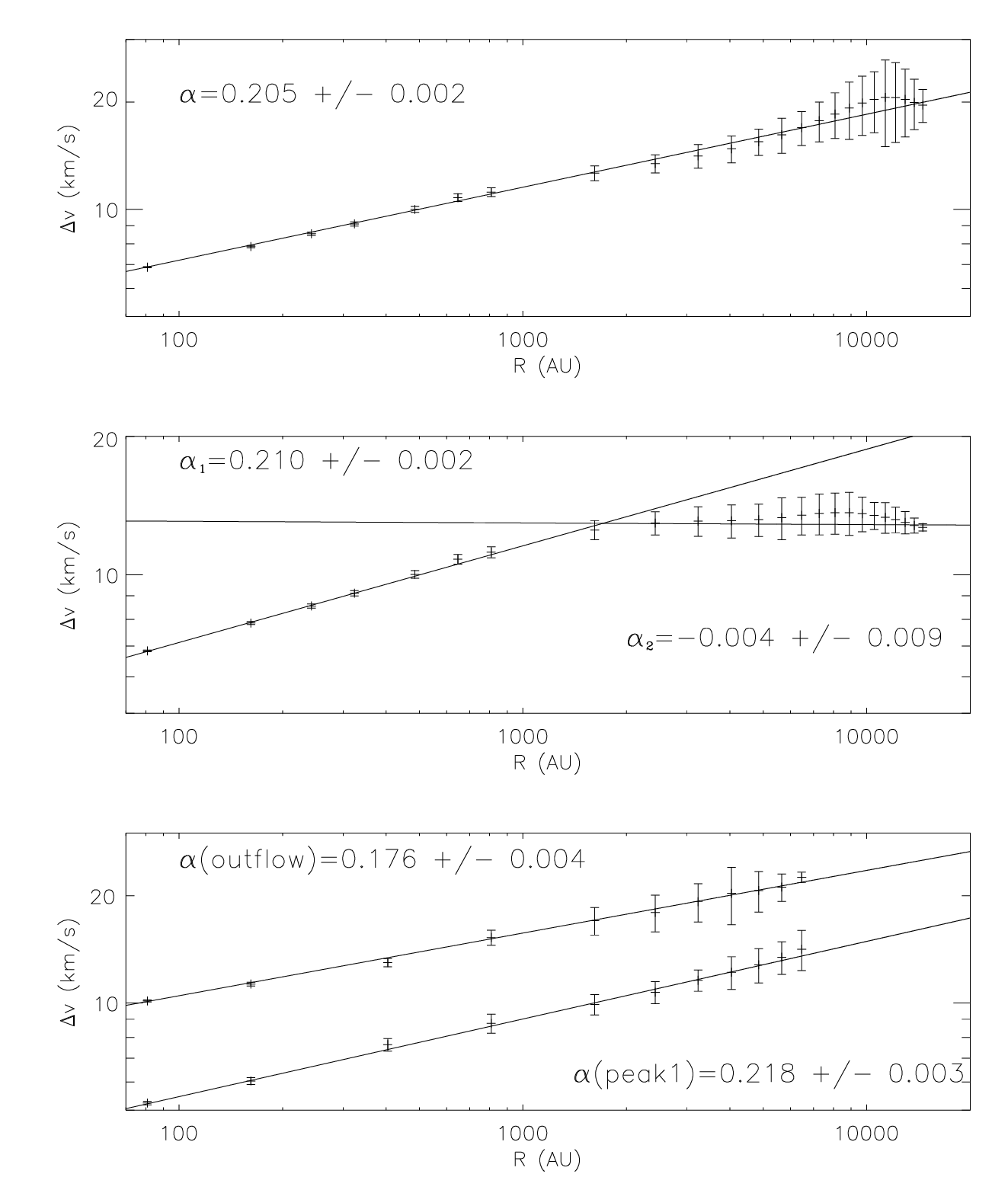}}
\caption{\small Larson size-linewidth relations, Eq.~(\ref{eqn:larson}). \textit{Top}: Velocity dispersion as a
function of size for the full velocity field. A power law fit of index
0.205 is overlaid. \textit{Centre}: Velocity dispersion for the filtered
velocity image. Power laws of index 0.210 and -0.004 are overlaid. \textit{Bottom}: Velocity dispersion for the outflow region (\textit{upper line}) and Peak 1 (\textit{lower line}) with power law fits overlaid.}
\label{fig:larson}
\end{figure}

The average
brightness weighted velocity
dispersion, $\Delta v_{obs}$, within regions of varying size, is estimated following the prescription of \cite{ossenkopf02}. We compute the
brightness weighted
probability distribution function of the peak velocities in a circular
region with a given radius and find the corresponding velocity
dispersion from a fit to a gaussian. This is repeated throughout the map and for a number of radii, R
(see Eq.~(\ref{eqn:larson})). For each radius the 
average velocity dispersion is calculated using intensity weighting. The relation between velocity dispersion and size is shown in
Fig.~\ref{fig:larson} for both the full velocity image (upper frame) and the
filtered velocity image (centre frame). Errors shown are the statistical standard
error on the mean.
Within the errors the relation for the full velocity field is
consistent with a single power law over more than two orders of
magnitude in R with an exponent $\alpha = 0.205 \pm 
0.002$, although there also appears to be some deviation above 8000 AU.
 The value of the exponent agrees with the average value of 0.21 $\pm$
 0.03 that 
\cite{caselli1995} found for massive cores in the Orion A and B clouds
at scales 0.03 - 1pc. It is striking that the Larson relationship appears both to hold in this very different regime and to have a very similar exponent as for massive cores, although we are dealing here with 
highly excited material.

The appearance of the region south-west of BN with fewer and more isolated clumps of
emission, suggests that the dominating
physical processes in this region are different from those in Peaks 1
and 2. In Peaks 1 and 2 the emission is more homogeneous and more
spatially concentrated. All clumps with
measurable radial velocities south-west of BN are
blue-shifted. \cite{nissen2005} provide clear evidence that these
objects form the IR counterpart of an outflow detected in radio
observations of SiO masers associated with a buried O-star within
OMC1 \citep{menten1995,doeleman1999,greenhill2004a,
greenhill2004b,shuping2004}. For this reason we refer to this zone as
the "outflow region".

 We have computed the size-velocity dispersion relation for the outflow
region and Peaks 1 and 2 separately, in each case for the full
velocity field (that is, without filtering of the data - see
Sect.~\ref{sec:filter}). The resulting size-velocity dispersion relationships,
shown for the outflow region and Peak 1, with Peak 2 omitted
for clarity, are shown in the third frame of
Fig.~\ref{fig:larson}. This illustrates that the structure adheres to
the Larson relationship in the outflow region south-west of BN but with a significantly lower slope and a somewhat higher proportionality 
parameter $\kappa$ in $\Delta v_{obs} = \kappa R^{\alpha}$ than for
Peaks 1 and 2. In the latter two regions $\kappa$'s and the exponents
$\alpha$ are the same within observational error. The lower value of
$\alpha$ associated with the outflow region supports the conclusion of
\cite{caselli1995} that lower exponents are characteristic of massive
star-forming regions.

The relation based on the filtered velocity image in the centre frame of Fig.~\ref{fig:larson}
shows that the velocity dispersion is not well
represented by a single power law. At radii smaller than $\sim$
1600~AU 
the relation follows a power law of index 0.210, essentially the same as for the full velocity field. This similarity of behaviour is expected since the filter applied to
remove the velocity gradient has a width of 14\arcsec $\sim$ 6400 AU
(Sect.~\ref{sec:filter}), and an equivalent radius of 3200AU, and thus should not affect smaller scales. In contrast to the full velocity field, the
velocity dispersion for the filtered data at larger scales is
constant within observational error as a function of size, in marked contradiction to the accepted form of
the Larson relationship, with an index at scales larger than 1600 AU of -0.004 $\pm$ 0.009. In
a turbulent medium, where the velocity distribution is expected to get broader
with increasing size of the region, this is unnatural and can only be
caused if some of the turbulent velocities have been artificially
removed. We conclude that by removing the large scale gradient we have
removed some of the turbulent energy. It follows therefore that the
large scale motions in OMC1, reported both here and by a number of other
authors \citep{scoville,sugai,chrysostomou1997,salas} should be seen as representing real scales of a 
turbulent cascade. By implication the turbulence is driven at large
scales and this suggests that turbulence on the scale of the
map could
be either the result of an energy cascade from still greater scales or
injection of 
turbulent energy by the large scale outflows associated with massive
star formation in the IRc2 region.

This in turn implies that the turbulence in the OMC1 region does not have a strong injection of energy at scales of less than 0.1 pc
and is thus not
primarily driven on small scales, for example by low mass protostellar
outflows. 
In support of this, the energy injected by a high mass
stellar outflow considerably exceeds that of the sum total of low mass outflows.
Elsewhere we have estimated \citep{nissen2005} that the mass outflow rate is
of the order of $>$ 10$^{-3}$ M$_{\odot}$/yr in the blue-shifted
outflow, SW of BN, with an average velocity of 18 kms$^{-1}$. Low mass
protostellar outflow rates are typically three orders of magnitude
lower \citep{richer}, with a
similar velocity. This simple argument on the basis of energetics supports
our conclusion that the injection of energy at the 0.1 pc scale of massive
stars outweighs on the global scale the energy input from low mass
stars. However, as we find in Sect.~\ref{sec:clump}, locally as opposed to globally, the character of the turbulence is strongly affected by low mass protostellar outflows.

The velocity
gradient, as identified above, should consequently not be removed prior to analysis. In the following
we thus mainly focus on the full velocity field containing velocities
on all scales, but for completeness we have also included analysis of
the filtered velocity image.

\subsection{Peak Velocity PDFs}
\label{sec:pdf_full}
The probability distribution function (PDF) of velocities is defined in terms of 3D velocity components as the
number of pixels at a certain velocity vs. velocity. Here we are restricted to a set of radial velocities and we estimate the PDF of these velocities sampled over the spatial region in the plane of the sky shown in
Fig.~\ref{fig:field}. This corresponds to 
the PDF of centroid velocities used in \cite{miesch1995,miesch1999,ossenkopf02}. An alternative
procedure is to use the line profile of an optically thin line as an
estimate of the PDF along the line-of-sight
\citep{falgarone1990,falgarone1991,falgarone1994,ossenkopf02}, taking advantage of the high velocity resolution of radio-observations. \cite{ossenkopf02} provides a comparison of the two methods.

The shape of the wings of the PDF is diagnostic of
intermittency, in which, at random scales and/or time intervals, energy
is dissipated as heat in the turbulent medium. At any scale the
removal of energy must involve those elements of the medium with the
most prevalent velocities. Thus it is the velocities at the centre
of the distribution which becomes most depleted. Thus the velocity
distribution loses its initial gaussian form, with the 
centre depressed and the wings relatively enhanced. 
Increasing degrees
of intermittency create a transition from gaussian ($\propto
\exp(-x^2)$) to 
exponential ($\propto \exp(-x)$) 
wings in a PDF.

Gaussian PDFs may be found in studies of decaying supersonic
turbulence \citep{ossenkopf02} and incompressible turbulence
\citep[e.g.][]{batchelor1953,jayesh1991}. Exponential PDFs can be
found in the literature of 
simulations, e.g. 
from inelastic collisions of clouds \citep{ricotti} and interactions
of shells \citep{chappell2001a}. Further relevant theoretical studies involve
the stretched exponential form ($\propto \exp(-a|x|^{\beta})$) where
$\beta$ is fractional \citep[e.g.][]{frisch1997,eggers1998}, where
$\beta =2$ reproduces the gaussian PDF and   
$\beta =1$ the exponential form. Fractional $\beta$ values can arise
from random processes whose overall effect accumulate in a
multiplicative manner (see Sect.~\ref{sec:2sf_full}).  
Other studies are concerned with wings of PDFs in the  
form of power laws for example arising from stellar winds or outflows
\citep{silk1995}.

Using line profile data for CO from various isotopes
\cite{falgarone1990,falgarone1991}, \cite{falgarone1994, 
  ossenkopf02} do not find simple gaussian behaviour for most 
observations. However,~\cite{falgarone1991} showed that most of the
PDFs could be represented by  
two gaussians, with the wing component being about 3 times broader
than the core component. The PDFs in the present work cannot be so
represented. Other work uses centroid velocities,
corresponding to the peak velocities used here.
For example, \cite{miesch1995}, \cite{miesch1999} found exponential
tails in most of their PDFs, while \cite{ossenkopf02} found that the PDFs of the Polaris
Flare could be reproduced by two gaussians as in
\cite{falgarone1991}. 

In the present work, the normalized PDF of peak velocities has been
calculated by binning 
the velocities in intervals  
of 1\kms. The PDF can be estimated by assigning the same weight to
every pixel or by weighting every velocity with the corresponding
brightness in that pixel. By testing both methods we found that the
chosen method of weighting does not influence the shape of the PDF
significantly. Brightness weighting is less influenced by
observational noise and is therefore used in the following analysis.

\begin{figure}
\resizebox{\hsize}{!}{\includegraphics{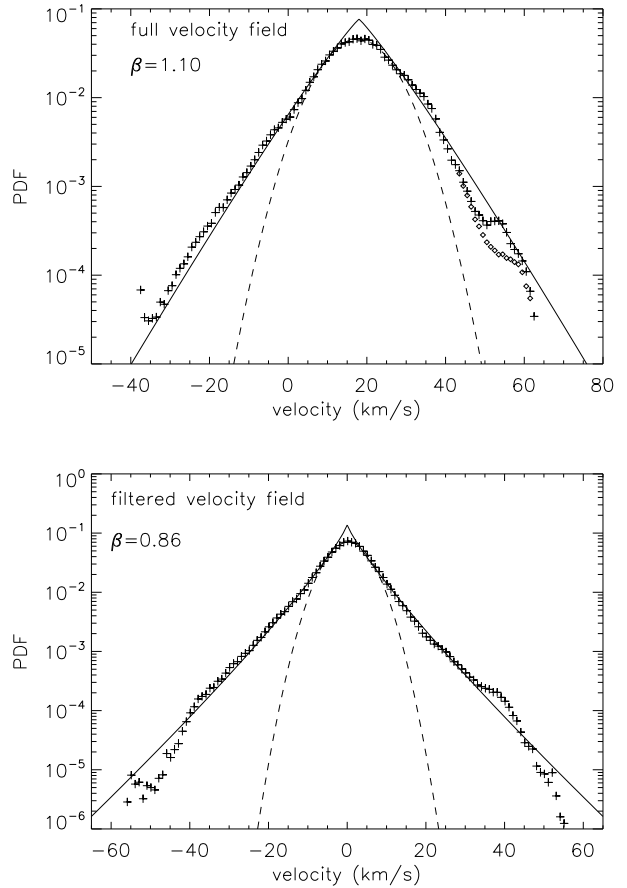}}
\caption{\small Probability distribution functions of velocities. Top: (+) PDF for the full velocity field. ($\Diamond$) PDF with removal of velocities associated with the clump in Fig.~\ref{fig:blob9}. Bottom: PDF for the filtered data (see text). Stretched exponentials with $\beta=1.10$
  and $\beta=0.86$ respectively
  are shown for comparison. The dashed lines are gaussian fits to the core.}
\label{fig:PDFcent}
\end{figure}

Fig. \ref{fig:PDFcent} shows the peak velocity PDF of the full
velocity field and of the filtered velocity image. The PDF of the filtered
velocity image is artificially
centered around 0~\kms ~through the action of filtering. The PDFs
are shown on a log-linear scale where a gaussian distribution would
form a parabola and an exponential a straight line. To test
the influence of uncertainties in the 
determination of velocities on the shape of the PDF, we have added
to each velocity in the original dataset a random velocity from a
gaussian distribution with standard deviation corresponding to the
uncertainty in the velocity of 
the pixel in question, as given by Eq.~(\ref{eqn:errors}). 
We created a
series of such velocity fields and calculated the PDF from each of
them. The PDFs of 
velocity fields so generated are effectively indistinguishable from
the PDF of the real data set, reflecting the fact that the large
number of pixels used in the 
PDF, 
$\sim$ 3.2$\cdot 10^6$, reduces the effect of the
observational noise to a negligible level. Again, since  every bin of the histogram 
is populated by a large number of pixels, the relative error,
$\sqrt{n}/n$, becomes very small, including taking account of the
effective 3x3 pixels resolution. Except for the very least populated 
bins the statistical error bars would be smaller than the symbols in
Fig.~\ref{fig:PDFcent} and have therefore been omitted.

The PDFs in Fig.~\ref{fig:PDFcent} are clearly not well represented by 
gaussians except
in the inner core, where gaussians are shown as dashed lines in
Fig.~\ref{fig:PDFcent}. Our data - for hot dense gas - 
show a different result to the data reported for much larger
scales in~\cite{falgarone1991} (for cool, diffuse gas and
  obtained using line profiles), where, as noted, a combination of
two gaussians fit the 
observations. The wings in the present work can be fitted by stretched exponentials (full lines). These lines are derived for the full velocity field data by fitting between +12 and -32 \kms and yield $\beta$=1.10$\pm$0.06. A similar fit for the filtered data yields $\beta$=0.86$\pm$0.05.

\begin{figure}
\resizebox{\hsize}{!}{\includegraphics{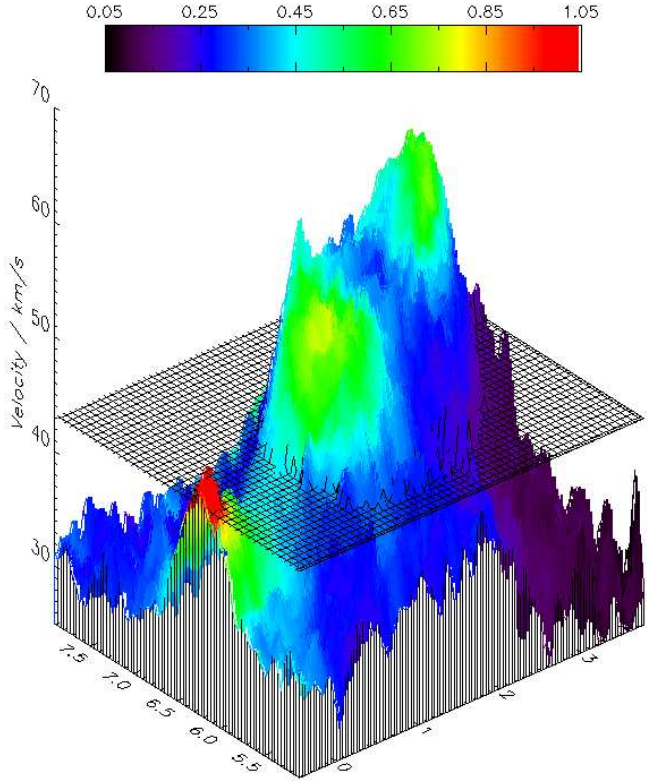}}
\caption{\small The brightly emitting clump that causes the hump in
  the red wing of the PDF (Fig.~\ref{fig:PDFcent}, + symbols). The xy
  plane is the plane of the sky, the vertical axis shows the radial
  velocity v$_{lsr}$. The plane at (x,y,43) shows the boundary of the region of
  pixels ignored in the PDF shown as ($\Diamond$) in Fig.~\ref{fig:PDFcent}. The centre of this clump is 6\farcs5 E, 1\farcs5 N of TCC0016. The colour scale refers to brightness in counts pixel$^{-1}$s$^{-1}$, see Sect.~\ref{sec:data}}
\label{fig:blob9}
\end{figure}

It is clear from Fig.~\ref{fig:PDFcent} that the PDFs are not
symmetric and that the red wing displays large deviations from a
smooth behaviour, especially at velocities of 50-60 \kms, where a hump
is seen in the PDF. By inspection of the velocity data we found that
the hump results from a single structure in the field. This structure
is the only clump of emission with velocities consistently larger than 50
\kms ~and is shown in Fig.~\ref{fig:blob9} in xy-velocity space. If we ignore
all pixels in this structure with velocities larger than 43 \kms, then
the PDF for the full field no longer contains a secondary hump at 55
\kms (diamonds in the upper frame in Fig.~\ref{fig:PDFcent}). This
suggests that the structure involved is an independent entity which
does not fit into the overall turbulent cascade. The nature of this
dense fast moving object remains mysterious.

The shape of the PDF(s) in Fig.~\ref{fig:PDFcent} can be further
quantified by the brightness weighted statistical moments, computed directly from the dataset as:
\begin{eqnarray}
mean&=&\mu=\frac{\sum_{map} I(x,y) v(x,y)}{\sum_{map} I(x,y)}\\
std.dev.&=&\sigma =\sqrt{\frac{\sum_{map}
    I(x,y)[v(x,y)-\mu]^2}{\sum_{map} I(x,y)}}\\ 
skewness&=&\frac{\sum_{map} I(x,y)[v(x,y)-\mu]^3}{\sigma^3 \sum_{map} I(x,y)}\\
kurtosis&=&\frac{\sum_{map} I(x,y)[v(x,y)-\mu]^4}{\sigma^4 \sum_{map} I(x,y)}
\end{eqnarray} 
where the summations represented by \textit{map} encompass all pixels in the data set.

\begin{table}
\caption{\small Moments and stretching exponents of the peak velocity
  PDF for the full velocity 
  field and the filtered velocity field}
\label{tab:PDFmom}
\begin{tabular}{ccccc}
\hline
\hline
Field&std-dev / \kms &skewness&kurtosis&$\beta$\\
\hline
full&10.3&-0.20&4.4&1.10\\
filtered&7.8&-0.042&6.4&0.86\\
\hline
\end{tabular}
\end{table}

The standard deviation quantifies the spread of the PDF, the skewness is a
measure of the asymmetry and the kurtosis characterizes the deviation
from a gaussian profile. A gaussian distribution has a kurtosis of 3
and larger values imply that the PDF has relatively more prominent
wings. An exponential distribution has a kurtosis of 6. The calculated
values for the PDF from our present data are listed in
Table~\ref{tab:PDFmom}, showing a departure from gaussian and from
pure 
exponential behaviour. We conclude that the data show clear evidence of
intermittency. There is also a skewness towards the blue. This may
arise through preferential obscuration of red-shifted flows, compared
to blue-shifted, in this dusty and partly obscured region.

\subsection{Variance and kurtosis of velocity differences}

\label{sec:2sf_full}

The PDFs presented above carry no spatial information. In order to
retain some of the spatial characteristics of the velocity field and
to quantify how velocities are spatially related within the medium, we
now construct the probability distribution function of velocity
differences between points separated by a certain distance in the
plane of the sky, the lag. The resulting distributions should provide
a more searching test of theoretical models than PDFs of
peak or centroid velocities.  

Velocity differences, $\Delta v=
v(\vec{r})-v(\vec{r}-\vec{\tau})$, are used in this analysis. Here both $\vec{r}$ and
$\vec{\tau}$, the lag, are two-dimensional space vectors. 
Previous
authors, using both observational data and theoretical models, the
latter for incompressible turbulence \citep{miesch1995,lis1998,miesch1999} have described how the shape of the PDF
changes with different lags. Specifically, PDFs were found to exhibit strong
non-gaussian forms at small lags.

For homogeneous, isotropic turbulence - as assumed here - the same
information can be obtained in a more compact form by investigating 
the structure functions as a function of lag magnitude,
L=$|\vec{\tau}|$. Structure functions of order p are defined as
\begin{eqnarray}
\label{eq:struc}
S_p (L) = <\mid v(\vec{r})-v(\vec{r}-\vec{\tau})\mid^p>=<\mid \Delta
v\mid^p>
\end{eqnarray} 
This is in
fact the traditional manner in which to characterise turbulence, as
Kolmogorov originally prescribed \citep{kolmogorov1941}. In the inertial
range of the turbulence the structure functions for Kolmogorov
turbulence (non-dissipative, incompressible) obey power laws, 
S$_p$(L)~$\sim$~L$^{\gamma(p)}$, and the exponents are called scaling
exponents.

Here we study the second order structure function, S$_2$(L), that is the variance, and
the kurtosis, S$_4$(L)/S$_2$(L)$^2$ 
as a function of lag magnitude, L.
In this manner information is obtained concerning the
evolution of the PDF as a function of lag
without the necessity for studying velocity difference PDFs
directly. We have included a weighting function for the
velocity differences, see Eqs.~(\ref{eq:variance}) and
\ref{eq:kurtosis}, and to avoid confusion with the traditional
definition of the structure functions (Eq.~(\ref{eq:struc})) we refer to
Eq.~(\ref{eq:variance}) as the variance function and to
Eq.~(\ref{eq:kurtosis}) as the kurtosis function.

Both the variance and the kurtosis functions are powerful tools 
when comparing observations with different models of turbulence.
For example, Kolmogorov turbulence predicts a power law variation of
the variance function (see Eq.~(\ref{eq:variance})) with lag magnitude, with a scaling exponent of 2/3, whereas~\cite{miesch1999} find values between 0.33 and
1.05 (see below). Numerical and analytical studies of driven
supersonic magnetohydrodynamic turbulence find a second order
structure function exponent of 0.74 \citep{boldyrev2002b}.

Constructing the variance and kurtosis functions involves combining
data from all pairs of pixels to 
obtain PDFs as a function of lag magnitude, L=$|\vec{\tau}|$.
The variance and kurtosis of the PDF of velocity differences are then 

\begin{eqnarray}
\label{eq:variance}
\sigma^2(L)&=&
\frac{\sum_{map}\sum_{|\vec{\tau}|=L}w(\vec{r})w(\vec{r}-\vec{\tau})(v(\vec{r})-v(\vec{r}-\vec{\tau}))^2}{\sum_{map}\sum_{|\vec{\tau}|=L}w(\vec{r})w(\vec{r}-\vec{\tau})}\nonumber\\
&~&
\end{eqnarray}

\begin{eqnarray}
\label{eq:kurtosis}
K(L)&=&
\frac{\sum_{map}\sum_{|\vec{\tau}|=L}w(\vec{r})w(\vec{r}-\vec{\tau})(v(\vec{r})-v(\vec{r}-\vec{\tau}))^4}{\sigma^4(L)\sum_{map}\sum_{|\vec{\tau}|=L}w(\vec{r})w(\vec{r}-\vec{\tau})}\nonumber\\
&~&
\end{eqnarray}
where the summations are performed over all pairs of pixels in the whole map
that fulfill the requirement $|\vec{\tau}|$=L. We have not used L $<$ 4 pixels (0.15 arcsec),
since this coincides with the spatial resolution of the data.

In the above, w is some weighting
function. An important issue is whether the velocity data should be brightness
weighted (w($\vec{r}$)=I($\vec{r}$) in counts per second) 
or equally weighted (w=1) in calculating the variance and kurtosis of the
velocity difference PDFs. In order to address this problem, we need further
to consider errors in the data. We have already removed data at the
2\% level as described in Sect.~\ref{sec:data}. We now examine whether
this cut-off is sufficiently low for the examination of variance and
kurtosis functions. We find that a 2\% cut-off is too low if we do not
use intensity weighting but that it is acceptable if intensity
weighting is included.

In order to investigate the effect of errors, we
use equal weighting and calculate the variance 
function for a   
number of cut-off values in brightness. 
All the variance functions can be 
approximated by power laws, but we find that the scaling exponent,
$\gamma$ in $\sigma^2 (L)\propto L^{\gamma}$, 
increases with increasing cut-off value before converging to a
constant value when the cut-off value reaches $\sim 9\%$ of
I$_{max}$. This indicates that  
pixels with brightness lower than this value contribute significantly
to the random noise and that they  
should be removed prior to analysis. This analysis emphasizes
that caution should always be exercised when equal
weighting is used since results may show dependence on the choice
of cut-off.

The brightness 
weighted variance function is expected to be less influenced by
noise. We find that 
the brightness weighted variance function including all pixels is
essentially the same as the equally 
weighted variance function with a brightness cut-off of 9\% of
I$_{max}$. Due to the lesser influence from noise we use the
brightness weighted variance and kurtosis functions in the following.

Errors in the brightness weighted variance function resulting from uncertainties in the
velocities have been calculated by using the error propagation
law and taking the 1$\sigma$ uncertainty on the velocity in each pixel
from Eq.~(\ref{eqn:errors}). Due to the large number of pixel pairs
that goes into the calculation of the variance function the relative error is
typically $10^{-3}$ and therefore negligible.

\begin{figure}
\resizebox{\hsize}{!}{\includegraphics{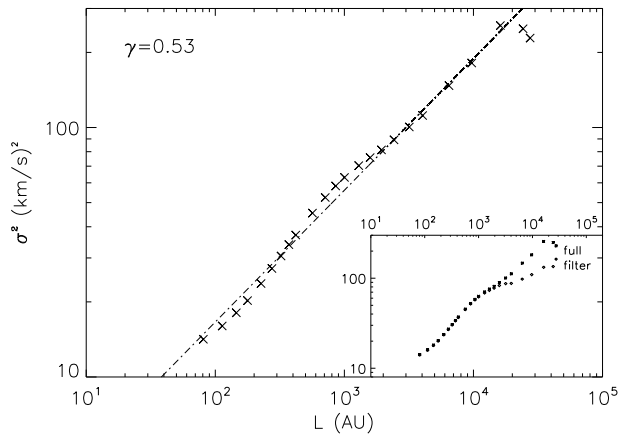}}
\caption{\small The variance function for the full
  velocity field. A power law form with exponent 0.53 is overlaid. The
  inset displays the variance function of the 
  filtered velocity image compared to the variance function of the full field.}
\label{fig:2sf}

\end{figure}

The variance function is shown in Fig.~\ref{fig:2sf} for the full
velocity field in a 
log-log plot.
The best fit of the variance function to a
power law is found to have  
$\gamma$=0.53$\pm$0.01 and this is also shown in
Fig.~\ref{fig:2sf}.
For comparison, \cite{miesch1999} obtained values of
$\gamma$ between 0.33 and 1.05 for several molecular clouds at scales
larger than 1.4$\cdot 10^3$ AU while \cite{ossenkopf02} obtained
$\gamma=0.94$ for the Polaris Flare at scales larger than 2000 AU. It
is noteworthy that our 
value of 0.53$\pm$0.01 is significantly lower than the 
Kolmogorov value of 0.67. We also note that the variance function is closely related to the Larson relationship such that the exponent of the variance function should be of the order of twice the Larson exponent. This is approximately satisfied here.

The variance function in Fig.~\ref{fig:2sf} in fact deviates significantly from a single power law, again underlining the
non-Kolmogorov nature of the gas dynamics. This is particularly
apparent through a positive deviation around 800 AU and a negative deviation around 100-200 AU. The curvature of the structure function may suggest the
presence of a multi-fractal medium. At all events, several power laws
operating in different ranges would appear necessary to fit the 
behaviour of the variance function.

The physical meaning attached to  $\sigma ^2$(L) in Fig.~\ref{fig:2sf} is
that it approximately represents all energies in eddies below any scale L
\citep{davidson2004}. The deviation from a power law in
Fig.~\ref{fig:2sf} below 
$\sim$~2000AU therefore represents more energy at lower scales and 
implies that there is an excess of material in
motion at 2000 AU and below. This scale suggests that the excess arises from outflow events associated with star
formation which result in energy injection at scales below 2000AU. This feature is apparent from the data presented in Paper 1 and in~\cite{nissen2005}. 
Conversely the material appears to suffer less velocity dispersion at
scales below 300AU. Thus material associated with these smaller scales appears to have dissipated some of its turbulent energy at larger scales. We speculate that these smaller less turbulent scales may be associated with protostellar nebulae or perhaps evaporating discs around protostars.
In earlier work, purely on the basis of spatially resolved
imaging \citep{vannier,lacombe2004}, a range of preferred scale sizes lying between 700 AU and 1100 AU was identified in Peak 2. In those cases and in the present case we suggest that the 
breaks in power laws are directly linked to scales associated with
star formation. 

The inset in
Fig.~\ref{fig:2sf} shows the effect of the removal of the large scale
gradient on the variance function. The variance function of the
filtered velocity image is identical 
to the variance function of the full field for scales smaller than
half the size of the 
filter ($\sim$ 3000 AU) above which the function flattens. It is clear that the
filter has no effect on very small scales 
and that the removal of the large scale velocity structure causes
the variance function to be systematically smaller on larger
scales. This effect of the 
filtering has also been noted by \cite{miesch1999} and is most likely an artefact due to the finite size of the map \citep{ossenkopf02}.

\begin{figure}
\resizebox{\hsize}{!}{\includegraphics{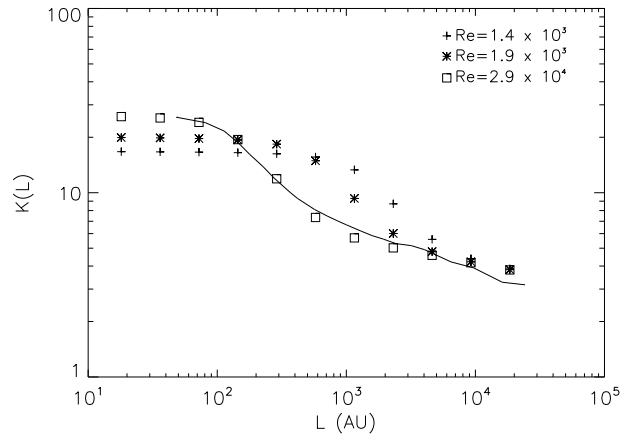}}
\caption{\small The kurtosis function, Eq.~(\ref{eq:kurtosis}), for the full
  velocity field (solid line). Theoretical predictions for 3 different Reynolds
  numbers from \cite{eggers1998} are
  shown for comparison.}
\label{fig:2sf_kurt}
\end{figure}

Turning now to the kurtosis function, the kurtosis is a measure of the
degree of correlation in the motion at a certain 
scale relative to overall motions seen in the map. 
A velocity field with no spatial correlations would display a
gaussian shape. As mentioned above, a gaussian has a kurtosis of 3. Thus
values of kurtosis exceeding 3 indicate the strength of velocity
correlation at specific lags. From simulations 
\cite{ossenkopf02} found that kurtosis values above 3 can only be
verified if the map has a scale considerably larger than the lag. 
As the lag approaches the size of the map, the value of the kurtosis
tends to approach the gaussian value of 3. This arises because the map
cannot of course contain larger scales than the size of the map itself. Thus a kurtosis value around 3 is 
always expected at scales approaching the size of the map and provides
no useful information on the velocity correlations at that scale.

Looking specifically at filamentary structure, \cite{lis1998} found
non-gaussian wings of velocity difference PDFs at small scales
evolving into gaussians at larger scales, without quoting values of
kurtosis. 
\cite{miesch1999} found kurtosis values between 10 and 30 at the
smallest lags (1.4$\cdot 10^3$ - 1.2$\cdot 10^4$ AU), with the
required gaussian value of 3 at large lags of the order of the map size (2$\cdot 10^6$ AU). Similar behaviour was found by \cite{ossenkopf02}.
The kurtosis function derived from our present data is shown in
Fig.~\ref{fig:2sf_kurt} as a continuous line. The kurtosis values approach 30 at small lags (100 AU)
and decrease gradually until reaching the value of 3 at
large lags. The shape of the
kurtosis function shows some similarity to those presented in
\cite{miesch1999}. However a much more striking comparison may be found 
with results of a model of multifractal processes in \cite{eggers1998}. In \cite{eggers1998} the kurtosis, called flatness in that work, is 
constant out to some lag beyond which the kurtosis decreases with
increasing lag. The shape of the kurtosis was found by
\cite{eggers1998} to depend on the Reynolds 
number, where R $>$1.1$\cdot 10^4$ marks a transition from a convex
shape to a concave shape. The shape remains essentially unchanged for
higher Reynolds numbers.

In order to compare with results in \cite{eggers1998} 
we make an estimate of the Reynolds number of the flows in OMC1.
The Reynolds number is given by Re = $ul/\nu$, where $u$ and $l$ are
the typical velocity and length scale, respectively, associated with
the gas flows. The dynamical viscosity, $\nu$, can be approximated by
$\nu 
\sim \lambda v_{th}$, where $\lambda$ is the mean-free-path of the
particles and $v_{th}$ is the rms thermal velocity
(kT/$\mu$m$_H$)$^{1/2}$ \citep{frisch1995}. Thus,
\begin{equation}
Re=\frac{ul}{\lambda v_{th}}=uln\sigma \sqrt{\frac{\mu m_H}{kT}}
\end{equation}
where the cross section, $\sigma$, for H$_2$-H$_2$ collisions is
2.7$\cdot 10^{-16}$ 
cm$^2$ \citep{atkins} and n is the number density. The shocked
regions observed 
here have a typical temperature of 2000-3000K and gas densities are
10$^6$ to 10$^7$ cm$^{-3}$.   
The velocity $u$ may be
approximated by the velocity dispersion in our observed region and $u$
is therefore given by twice the   
standard deviation of the PDF of peak velocities (Table~\ref{tab:PDFmom}).
The corresponding $l$ is the size of the region $\sim$ 3$\cdot 10^4$ AU. 
Using these values, we obtain a Reynolds number of
Re~$\sim 3 \cdot 10^{9}$. The significance of this value is only that it is
very large and greatly exceeds the critical value in \cite{eggers1998} for which convex behaviour of the kurtosis function vs lag becomes concave.  
If we were to include magnetic effects, the
magnetic Reynolds number might be substantially higher.  

The \cite{eggers1998} description of turbulent processes is ad hoc in the
sense that it does not result from detailed high resolution three
dimensional
MHD simulations. Instead some reasonable assumptions are made concerning
the
cascade of turbulent energy, assuming it to be a stochastic process,
isotropic and homogeneous in space, with certain additional properties
which are briefly summarized below.
Once one assumes that the turbulence is isotropic and homogeneous, the turbulence in the Fourier domain is represented by a spectrum which
is a function of a single variable $r$, corresponding to a size of
turbulent eddies.
The approach of \cite{eggers1998} is to impose a recipe for the cascade
of energy from larger to smaller scales. The recipe states that energy
at a given scale $r$ is transferred {\em only} to a scale $r/2$.
The {\em amount} of energy transferred, or more precisely the ratio
$s$ of the velocity amplitudes between scale $r$ and $r/2$ in equilibrium,
is given by a probability distribution~:
\begin{eqnarray}
p(s) &&= p \delta(s-s_1) + (1-p) \delta (s-s_2)
\label{pp}
\end{eqnarray}
with
\begin{displaymath}
p = 0.688,\qquad s_1 = 0.699,\qquad s_2 = 0.947
\end{displaymath}
where the probability $p$ and parameters
$s_1$ and $s_2$ shown above were obtained by fitting to reproduce
laboratory experimental data.

Equation~(\ref{pp}) states that the probability $p$ is high ($\sim 2/3$) for the
ratio of velocities to be relatively small ($s=s_1$), which
corresponds to a low efficiency of 
energy cascade from one scale to the next. However there remains a probability of $\sim 1/3$ that there is efficient cascading, $s=s_2$.
Starting from some chosen outer scale, this recipe allows for a very quick
calculation of the energy at any given smaller scale by multiplying
the probabilities for every cascade step lying between these two
scales and producing the appropriate PDF semi-analytically. Hence the
term multiplicative turbulence. The
viscous cut-off of the turbulence is implemented through stopping the
further cascading of energy if the Reynolds number at a given scale
falls below a set value. The velocity fluctuation spectrum is replaced below that scale by an exponentially decaying tail.

We show in Fig.~\ref{fig:2sf_kurt} a comparison between our observed
kurtosis and that presented in \cite{eggers1998} for Reynolds numbers
between $1.4\cdot10^3$ and $2.9\cdot10^4$, the highest which \cite{eggers1998} show, noting once more that the form of the kurtosis function appears to be independent of
Reynolds number above $\sim ~10^4$. The lower Reynolds number cases
are shown purely for comparison. The similarity of the kurtosis
function between that for our data and that of \cite{eggers1998}  
suggests that the model of a multifractal, multiplicative turbulent medium on which \cite{eggers1998} is based may capture some of the physics of the turbulence in
the hot component of the gas in OMC1.

\section{Analysis of individual clumps}
\label{sec:clump}

The high spatial resolution of the data allows the analysis of
individual clumps of shocked gas with the same statistical
methods as above. First, the dimensions of individual clumps must be
defined and the following algorithm is used to identify the extent of
any clump. The data are smoothed by a 9 by 9 pixel boxcar, in order
that random fluctuations become unimportant. A clump is defined as a
region which encompasses all pixels surrounding a local brightness
maximum where the 
emission is situated on a continuously decreasing slope from the maximum
brightness. Thus when we move in any direction
starting from the pixel of maximum brightness and encounter a pixel with
higher brightness than the preceding pixel, we define this as the
boundary of the clump in this direction. As a further constraint we
only use pixels with a brightness larger than 20\% of the local
brightness maximum. This avoids clumps becoming unrealistically large
in the outer regions with 
low brightness and few brightness maxima. The 20\% restriction has no
effect in bright congested regions such as Peak 1 and 2. In passing, we note that we cannot rule out that some clumps so identified are the
result of chance 
superpositions of two or more isolated clumps in the same 
line-of-sight. We have ignored this possibility.

With the above definition, we have delineated 170 clumps. These are in
most part essentially the same features as analyzed in
\cite{nissen2005}. The number of 
pixels in each clump ranges from 841 to 18026, which corresponds to
sizes of clumps between  approximately 500 and 2200 AU. Due to the small size
of the clumps, the earlier discussion concerning the removal of the
large scale gradient is irrelevant here, since filtering does not
affect the velocity field over such restricted regions.

\subsection{PDFs of clumps}
\label{sec:pdfclump}
\begin{figure}
\resizebox{\hsize}{!}{\includegraphics{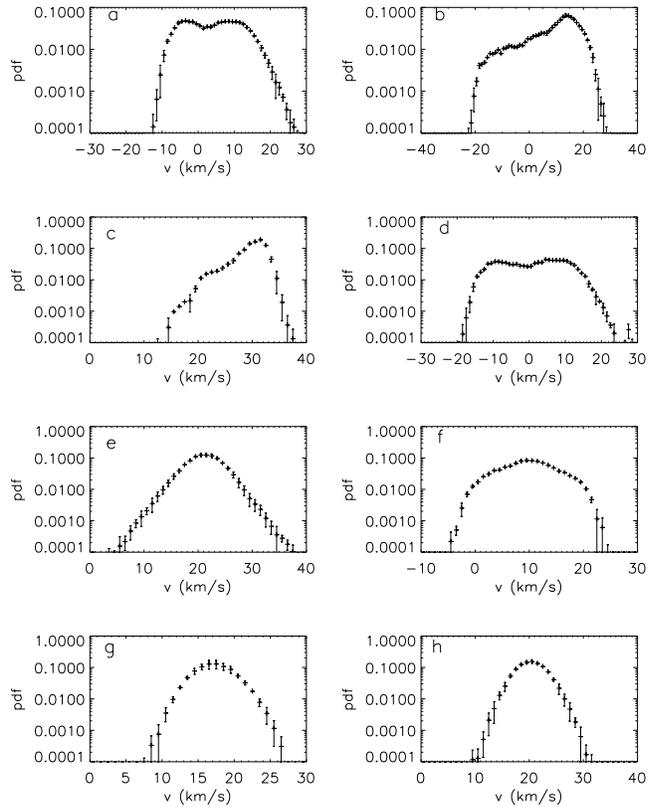}}
\caption{\small PDFs of eight representative clumps. a), b), c) and d) are
  multi-modal, e) and f) are stretched exponential, g) and h) are
  gaussian. Positions are shown in Figs.~\ref{fig:pdf_form},
  \ref{fig:clump_l}, \ref{fig:clump_s}}
\label{fig:pdf_clump}
\end{figure}

The PDFs of the 170 individual clumps have
been calculated. The sensitivity of the shape of the PDF
to uncertainties in the velocities has been investigated using
simulations in the same
way as in Sect.~\ref{sec:pdf_full}. Errors vary from clump to clump
but do not influence the 
general shape of the PDF in any of the clumps. Eight representative
PDFs with error estimates are shown in
Fig.~\ref{fig:pdf_clump}. Error bars shown represent the values spanned
by the observed data and the simulated data. The
PDFs for the clumps take on many different 
shapes. For
most clumps the shape is complex and appears bi- or multimodal. Bi- or
multimodal PDFs, such as shown in Figs.~\ref{fig:pdf_clump}a, b, c and d, can result from bipolar outflows from one or multiple protostellar
objects. Most stars form in binary or multiple systems and the jets
from these are found to be episodic or pulsed \citep[][and references therein]{larson2003}. 39 
clumps have clear stretched exponential wings. Two such clumps are shown
in Figs.~\ref{fig:pdf_clump}e,f. 27 clumps have
PDFs that can be fitted by a single gaussian. Two examples are shown
in Figs.~\ref{fig:pdf_clump}g,h.

\begin{figure*}
\centering
\includegraphics[width=13.2cm]{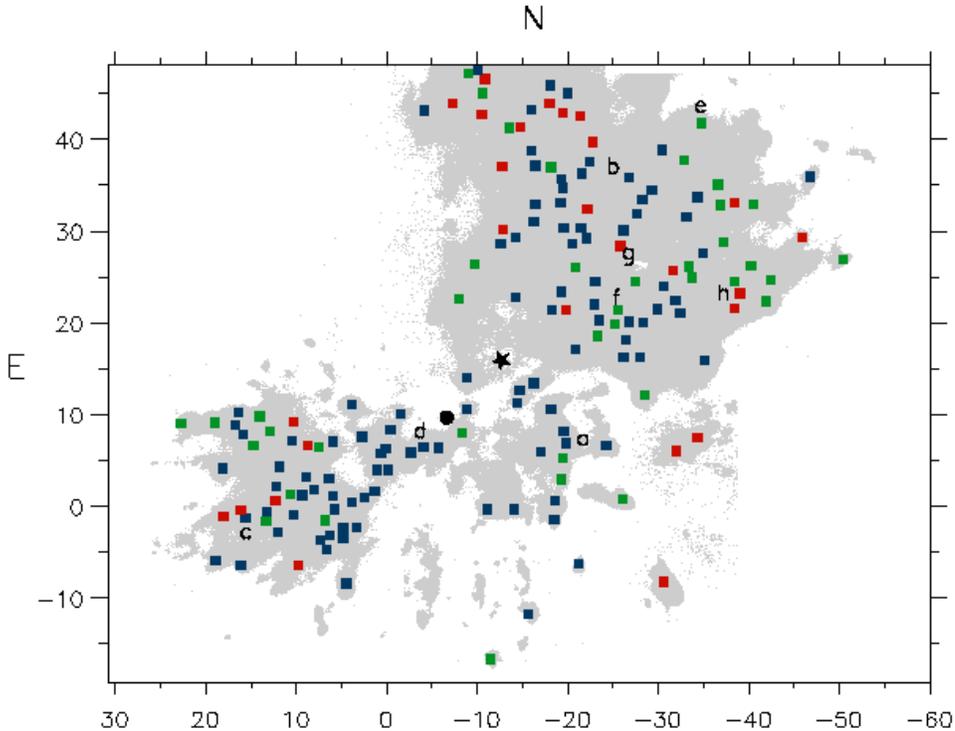}
\caption{\small Position of clumps with bi- or multimodal PDF (blue),
  stretched exponential PDF (green) and gaussian PDF (red). The underlying grey background represents the spatial extent of velocity integrated H$_2$ emission at 2.121 $\mu$m with brightness greater than $\sim$~2.4 x 10$^{-6}$Wm$^{-2}$sr$^{-1}$. The position of BN is marked with a
  star, IRc2 with a circle.  a,b,c,d,e,f,g,h mark clumps
  shown in Figs.~\ref{fig:pdf_clump}, \ref{fig:2sf_clump}.}
\label{fig:pdf_form}
\end{figure*}

The spatial distribution of clumps with gaussian PDFs, stretched exponential
PDFs and bi- or multimodal PDFs is shown in Fig.~\ref{fig:pdf_form}.
Clumps with gaussian PDFs are found in all regions of the observed
field, except in the central region around BN-IRc2, with essentially
the same distribution as clumps with non-gaussian 
PDFs. That is, 
there is no tendency for clumps with gaussian PDFs, for example, to group together in small
areas. Thus it seems that the turbulence dominating the BN-KL nebula
can simultaneously produce both gaussian and non-gaussian PDFs in clumps
with sizes 
of $\sim 10^3$ AU, and that the clumps with gaussian and non-gaussian PDFs are
intermingled. However there appears to be some mechanism that inhibits the
formation of clumps with gaussian PDFs in the outflow region around
BN-IRc2.     

In connection with the above, gaussian PDFs would of course arise if the velocity data had a significant noise contribution. However it turns out that 
this is not the case save perhaps in 3 of the 27 clumps mentioned above. 
For six of the 27 clumps with gaussian 
PDFs, the 1$\sigma$ uncertainty on the velocities
(Eq.~(\ref{eqn:errors})) exceeds the
half-width (one standard deviation) of the gaussian PDF in 1\% or more of the  
total number of pixels. The numbers of pixels are 1\%, 4\%, 5\%, 28\%, 32\% and 54\% of the totals. In the latter three of these six clumps the velocity
distribution could be 
dominated by uncertainties in the velocities and these data may
therefore be spurious, in the sense that the gaussian character may be
over-emphasized by noise.

\subsection{PDFs of velocity differences: variance functions for clumps}

\begin{figure}[h!]
\resizebox{\hsize}{!}{\includegraphics{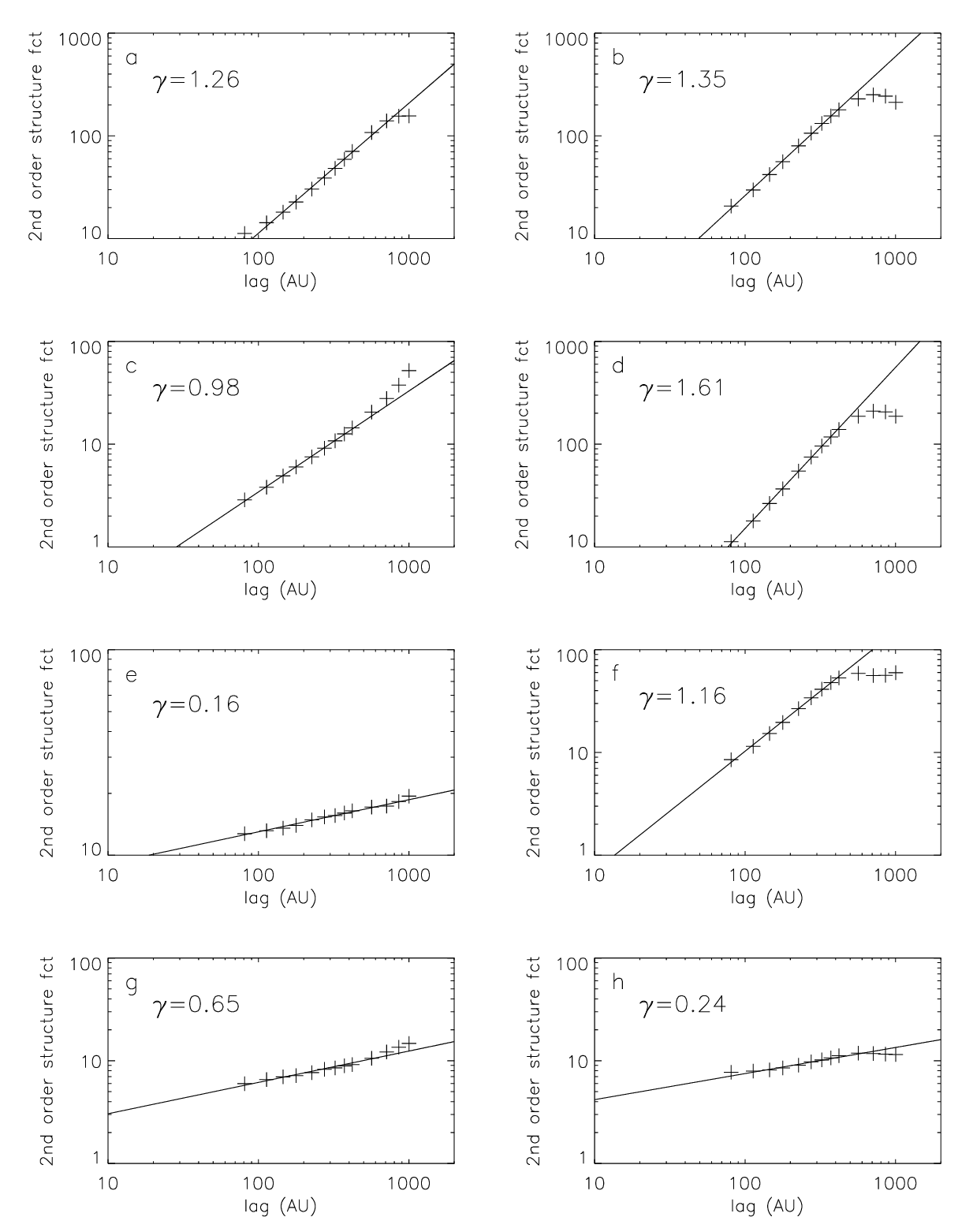}}
\caption{\small The variance function for the eight 
  individual clumps shown in Fig.~\ref{fig:pdf_clump}. Power law fits are overlaid and the value of the
  exponent is given.}
\label{fig:2sf_clump}
\end{figure}

The variance function of the
velocity field has also been
calculated for each clump, using Eq.~(\ref{eq:variance}). Most of the variance functions
appear to be well approximated 
by power laws with a few exceptions, but the value of the scaling exponent $\gamma$ varies between essentially zero, that is, a flat distribution, and 1.61. The variance functions are 
shown in Fig.~\ref{fig:2sf_clump} for the eight representative clumps shown in Fig.~\ref{fig:pdf_clump}. 

The
errors in the variance functions have been calculated in the same way as for the full
region in Sect~\ref{sec:2sf_full}. The magnitude of the errors
depend on the brightness (see Eq.~(\ref{eqn:errors})) and the physical
size of the clump, where size is the dominating parameter. Thus the smaller clumps have the larger errors, but even the smallest clumps have relative errors of no more than 10\%. Two of the clumps in
Fig.~\ref{fig:2sf_clump} have relative errors of 10\% which is of the
same order as the size of the symbols in Fig.~\ref{fig:2sf_clump}. Thus the errors have
been omitted in Fig.~\ref{fig:2sf_clump}. Errors in values of $\gamma$
arise from lack of precision of the power law fit, rather than from
random errors in the variance. Typical errors in $\gamma$ are $\pm$0.01.  Clumps with
gaussian PDFs show $\gamma$ = 0.0 - 0.97 including or excluding the 3
clumps with significant noise contributions (see Sect.~\ref{sec:pdfclump}).
Clumps with
non-gaussian PDFs cover the whole range of $\gamma$ = 0.0 - 1.61.

\begin{figure*}
\centering
\includegraphics[width=12cm]{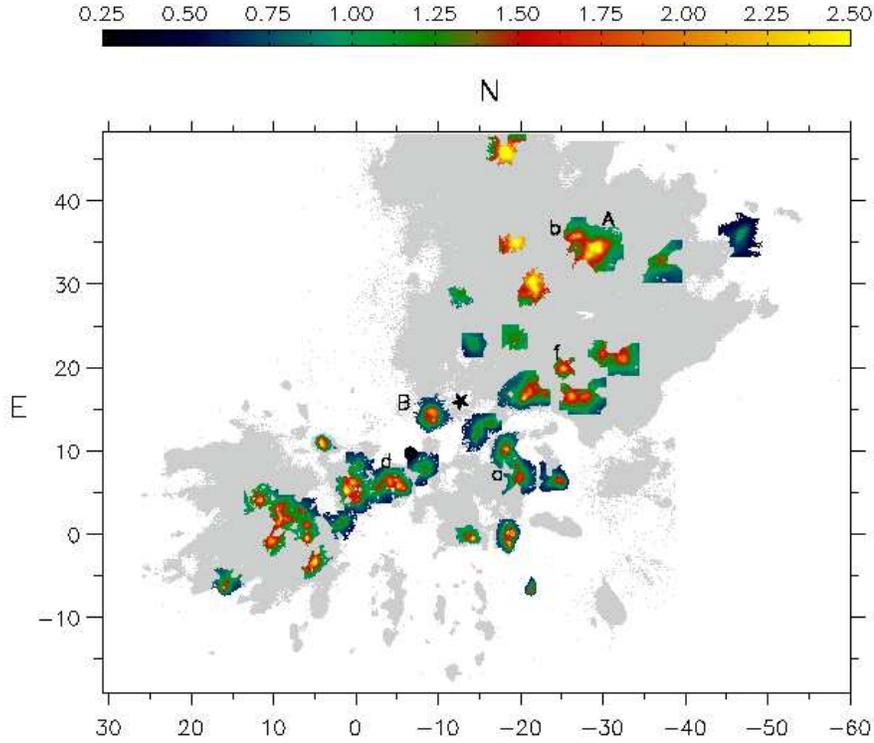}
\caption{\small Clumps with the variance function scaling exponent $\gamma > 1.0$. As in Fig.~\ref{fig:pdf_form}, the underlying grey background represents the spatial extent of velocity integrated H$_2$ emission at 2.121 $\mu$m with brightness greater than $\sim$~2.4 x 10$^{-6}$Wm$^{-2}$sr$^{-1}$. The position of BN is marked with a
  star, IRc2 with a circle. Colours represent brightness in counts per
  pixel per second in the clumps analyzed in sect. \ref{sec:clump}. A
  and B mark sites of protostellar candidates. a,b,d,f mark clumps
  shown in Figs.~\ref{fig:pdf_clump}, \ref{fig:2sf_clump}.}
\label{fig:clump_l}
\end{figure*}

\begin{figure*}
\centering
\includegraphics[width=12cm]{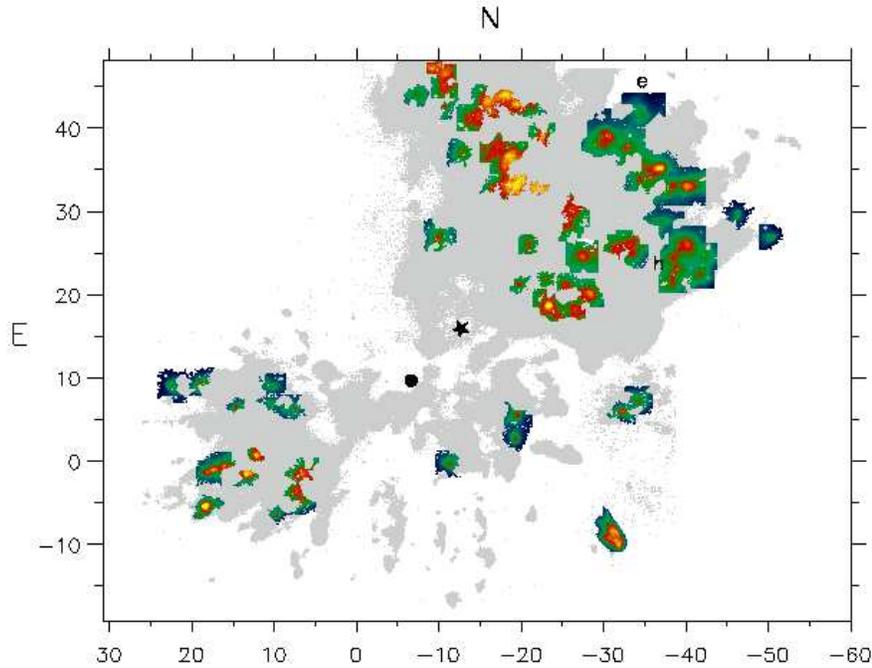}
\caption{\small Clumps with the variance function scaling exponent $\gamma < 0.67$. Grey background as in Fig.\ref{fig:clump_l}. The position of BN is marked with a
  star, IRc2 with a circle. Colours represent brightness in counts per
  pixel per second in the clumps analyzed in Sect. \ref{sec:clump}. e,h
mark clumps shown in Figs.~\ref{fig:pdf_clump}, \ref{fig:2sf_clump}. }
\label{fig:clump_s}
\end{figure*}

A high value of $\gamma$ represents more ordered structure in the velocity
field and large velocity gradients within the clump. This
suggests the presence of ordered shock structure. Figure~\ref{fig:clump_l} shows the spatial distribution of clumps with $\gamma > 1.0$, that is, clumps with the highest degree of order in the velocity field. The
less ordered clumps with $\gamma < 0.67$, the Kolmogorov
value, are shown in Fig.~\ref{fig:clump_s}. Examination of Figs.~\ref{fig:clump_l} and  \ref{fig:clump_s} shows that the
value of $\gamma$ does not correlate with brightness, that is, high $\gamma$
can occur equally for strong and weak emission clumps. Furthermore, in
Peaks 1 and 2 - but not in the central region around BN-IRc2, see
below - clumps with 
high $\gamma$ may be found mixed with clumps with low $\gamma$ in the same
spatial region. A theory of star formation including turbulence,
self-gravity, bipolar outflows, etc. should thus be able to reproduce
such a range of exponents of the variance function
within a limited physical region.

The majority of clumps with $\gamma$ less than the Kolmogorov value of
0.67  resides in Peaks 1 and 2 as seen in  Figs.~\ref{fig:clump_l} and
\ref{fig:clump_s}. The major part of these corresponds to clumps with measured velocities which display no clear
structure and whose relative velocities within any clump do not exceed 5
kms$^{-1}$ \citep{nissen2005}, accordingly giving low values of the exponent. We suggest that some of these regions may arise from 
photodissociation regions (PDRs) involving an irregular surface
containing a variety of lines of sight, illuminated by
$\theta$$^1$Ori-C, in a model essentially as described in \cite{field1994} for the PDR NGC2023. This would be
consistent with 
a region lacking regular structure. However some low $\gamma$ clumps
are very bright and cannot be reconciled with a PDR excitation model. These are likely to be shocks travelling across the line of sight.

Many clumps with high $\gamma$ (Fig.~\ref{fig:clump_l}) are situated
in the vicinity of 
the BN-IRc2 complex in the central region. Moving away from
BN-IRc2, their occurrence becomes progressively less. Some of the high
$\gamma$ clumps possess morphologies which resemble bow 
shocks 
(e.g. at (-18\arcsec, 0\arcsec) and (-21\arcsec,-7\arcsec), see
Fig.~\ref{fig:clump_l}) 
and are part of the blue-shifted lobe of a bipolar outflow from
radio source I, itself associated with a massive
O-star
\citep{menten1995,doeleman1999,greenhill2004a,greenhill2004b,shuping2004,nissen2005}.
Other high $\gamma$ clumps 
correspond to protostellar candidates identified in
Paper I and arise from outflows created internally in the
clumps e.g. at 
(-30\arcsec,35\arcsec) and (-10\arcsec,15\arcsec), marked A and B in 
Fig.~\ref{fig:clump_l}.

In contrast to regions of high $\gamma$, regions of low $\gamma$ are
notably absent around BN-IRc2 and in the vicinity of source I, and
tend to congregate where regions of high $\gamma$ 
are scarce. Our view of turbulent motion, based on models of turbulence,
like that for example of \cite{eggers1998}, deals only with spatial
scales and not directions. Thus we tend to consider only isotropic
turbulence. However in the central region between Peaks 1 and 2, and also
to the south-west of BN, the outflow from source I imposes a strong constraint on the radial motion of clumps in that region, where as mentioned clumps show a strong blue-shift. Motion
is evidently not isotropic here. We suggest that high values of
$\gamma$ arise in this region because effects of turbulence in the radial
direction are effectively swamped in the variance function
by large blue-shifted motions and the turbulence tends to 2D rather than
isotropic 3D. As we move away from the region of the blue-shifted outflow,
low $\gamma$ clumps are once more encountered (see
Fig.~\ref{fig:clump_s}). 
It is an interesting challenge to theory to test if
imposition of a strong outflow has the effects that we observe here on
values of $\gamma$, as a check on our interpretation. Models should
also demonstrate the absence of gaussian PDFs.

We now consider high $\gamma$ regions outside the blue-shifted outflow
zone, that is, those which are well-mixed with regions of low $\gamma$.
The formation and evolution of protostellar objects involve periods where 
a high degree of order in the velocity field surrounding the protostar
is expected. A bipolar outflow encountering the circumstellar envelope
would for example shock-excite the gas while creating an organized
velocity field, with accompanying high $\gamma$. The detection of clumps with
high values of $\gamma$ could therefore 
potentially allow an independent method of identifying early
protostellar objects in the Class 0/1 phase. This method may prove of value with the advent of very high spatial resolution radio maps with the Atacama Large Millimetre Array.

\section{Conclusions}
\label{sec:conc}

The fraction of gas studied in the present work is highly excited and very dense and quite distinct from gas whose statistical properties have been studied in earlier work. The latter refers to much larger scales and to much cooler, lower density and relatively quiescent gas. Nevertheless we have found that the hot dense gas in OMC1 nicely follows the general trends observed in earlier studies, down to the smallest, densest scales investigated here.

Our major conclusions from the statistical analysis of flows of H$_2$ in OMC1 may be outlined as follows:

1. The size-linewidth relation of \cite{larson1981}
is recovered at the small scales investigated here. The scaling
exponent is $\alpha = 0.205 \pm 0.002$, which agrees with the
average value of 0.21 $\pm$ 0.03 that \cite{caselli1995} found in
Orion at scales 
0.03 - 1pc, using CO as an indicator (and which did not in fact include OMC1). 

2. We find that the outflow, shown elsewhere \citep{nissen2005} to originate from a young and deeply buried
O-star, between Peaks 1 and 2, generates structure which also follows
the Larson relationship. Thus quite different environments preserve the Larson
relationship. The velocity dispersion is however higher for any given
size of object than elsewhere in the field and the exponent is significantly smaller than in the full region.  

3. If we use velocity filtered data, removing the largest scales which appear as an apparent gradient of velocity, the Larson relationship breaks down. We conclude
that the large scale motions are an inherent part of the turbulent
cascade and should therefore not be removed from the data. Large scale injection of turbulent energy is the
dominant process and outflows from massive star(s) in the IRc2 complex
may contribute a substantial part of the driving of the turbulence at the
scale of 0.1 pc. 

4. The probability distribution function (PDF) of velocities is best
fitted by an exponential or a weakly stretched exponential and departs strongly from gaussian. This suggests that
the turbulence in the region studied here is characterized by intermittency.
%and the variance function of the velocity differences, the second
%order structure function, shows that a single fractal model is too
%simple. 

5. The variance function of the velocity differences is not well represented by a single power
law. A multifractal model is implied. The best fit scaling
component is 0.53$\pm$0.01, significantly lower than the Kolmogorov
value of 0.67, underlining the non-Kolmogorov nature of the turbulence.

6. The behaviour of the variance function shows that there is a preferred scale size in the medium below 2000 AU reflecting the presence of protostellar
zones of this dimension and below. There is also clear evidence of
less turbulent structure below~$\sim$300AU, perhaps representing a
population of protostellar disks which have expelled part of their
turbulent energy. 

7. As further evidence of the multifractal nature of the medium, the
kurtosis function of the velocity 
differences closely resembles that of the multifractal
model of \cite{eggers1998} for high Reynolds numbers. 

8. Analysis of 170 individual clumps with sizes between 500 and 2200 AU
opens a window on a regime of scales that has never before been explored
using statistical techniques. Studies at these scales reveal
considerable diversity of velocity PDFs, with some gaussian, some showing evidence of intermittency and many with complex structure, reflecting multiple outflows. Variance functions are approximately power laws, with
exponents varying between zero and 1.61. There is no spatial association between high and low exponent clumps nor is there any association between gaussian PDFs and high or low exponents. 

9. To emphasize the last point, clumps with a variety of forms of the PDF and different values of the scaling
exponent of the variance function are found in the same spatial region and in
the whole region covered by the observations.

10. An outflow region associated with a deeply embedded O-star between
Peaks 1 and 2 shows markedly different statistical characteristics
from Peaks 1 and 2. Clumps in the vicinity of the BN-IRc2 complex
typically show high scaling exponents in the variance
function. This may be due to the action of the energetic outflow
imposed on intrinsic turbulent motion within individual clumps. At all
events high and low exponent clumps are spatially segregated in this
zone, with low exponent clumps found only at the edges. 

11. The outflow region apart, our results suggest that analysis using variance functions could be a useful way in which to establish the presence of early star-forming regions. 

The above results constitute a challenge for numerical simulations of turbulence in star forming regions. To our knowledge the
resolution of any present numerical simulation is substantially lower than the
resolution of these observations, but advances in computer technology
will soon allow simulations to reach the scales encountered here. A
self-consistent theory of star formation including self-gravity and
MHD turbulence should be able to reproduce the features outlined in
this paper. Thus the size-linewidth relation, the probability distribution
function of peak velocities and the variance and the kurtosis of
velocity differences as a function of lag form a dataset which models of star-forming regions
should aim to reproduce.

\begin{acknowledgements}
The authors would like to thank Dr. Wang for supplying us with
numerical values for inclusion in Fig.~\ref{fig:2sf_kurt}. DF and MG
would like to acknowledge the support of the Aarhus Centre 
for Atomic Physics (ACAP), funded by the Danish Basic Research
Foundation and the Instrument Centre for Danish
Astrophysics (IDA), funded by the Danish Natural Science Research
Council. DF would also like to thank the Observatoire Paris-Meudon
for support during the period in which this work was performed. JLL
would like to acknowledge the support of the PCMI National Program
funded by the CNRS. We would also like to thank the Directors and
Staff of the 
CFHT for making possible the observations reported in this paper. We
should also like to acknowledge the helpful comments made by the
referee. 
\end{acknowledgements}

\bibliographystyle{aa}
\bibliography{bibliography}

\end{document}